\documentclass[superscriptaddress,twocolumn,showkeys]{revtex4}
\usepackage{textcomp}
\usepackage{amsmath}
\usepackage{amssymb}
\usepackage{graphicx}
\usepackage{epstopdf}
\usepackage{float}
\usepackage{booktabs}
\usepackage{dcolumn}
\usepackage{bm}

\begin{document}

\title{On the PeV knee of cosmic rays spectrum and TeV cutoff of electron spectrum}

\author{Chao Jin}
\email{jinchao@ihep.ac.cn}
\affiliation{Key Laboratory of Particle Astrophysics, Institute of High Energy Physics, Chinese Academy of Sciences,Beijing 100049,China}
\affiliation{University of Chinese Academy of Sciences, 19 A Yuquan Rd, Shijingshan District, Beijing 100049, P.R.China}
\author{Wei Liu}
\email{liuwei@ihep.ac.cn}
\affiliation{Key Laboratory of Particle Astrophysics, Institute of High Energy Physics, Chinese Academy of Sciences,Beijing 100049,China}
\author{Hong-Bo Hu}
\email{huhb@ihep.ac.cn}
\affiliation{Key Laboratory of Particle Astrophysics, Institute of High Energy Physics, Chinese Academy of Sciences,Beijing 100049,China}
\affiliation{University of Chinese Academy of Sciences, 19 A Yuquan Rd, Shijingshan District, Beijing 100049, P.R.China}
\author{Yi-Qing Guo}
\email{guoyq@ihep.ac.cn}
\affiliation{Key Laboratory of Particle Astrophysics, Institute of High Energy Physics, Chinese Academy of Sciences,Beijing 100049,China}

\date{\today}

\begin{abstract}
Spectra of Cosmic Rays (CRs), and particularly their features, contain a lot of important information about the astro-particle physics and the fundamental physics. Realizing that the 4 PeV knee of CR nuclei and the 1 TeV spectral cutoff of the electron share almost a same Lorentz factor, we propose that CRs experience a threshold interaction with a new light particle X abundant in the Galaxy. The interaction $CR \ + \ X \ \longrightarrow \ CR \ + \ X^{\prime}$ can take place when the effective energy is sufficient to convert it into another heavier unknown particle $X^{\prime}$ (as  a representative to all possible threshold inelastic interactions). Under this scenario, we can reproduce the spectral break for both the nuclei and the electron, and predict a zig-zag spectrum for them. Given that there are great uncertainties in experiments, our model accommodates a wide mass range of the $X$ from ultra low value to around 1 eV.
\end{abstract}
\maketitle

\section{Introduction}
The origin of the CR knee around 4 PeV has remained a puzzle since its discovery \citep{1959JETP...35....8K}. In general, three types of mechanism have been proposed, including the acceleration limit of the CR sources \citep{1999JETP...89..391B, 2002PhRvD..66h3004K, 2001A&A...369..269B, 2003A&A...409..799S}, the diffusion process of the galactic CRs \citep{1993A&A...268..726P, 1998A&A...330..389W}, or the energy loss with the radiation background at the source location \citep{2009ApJ...700L.170H, 2010SCPMA..53..842W}.
\par
In the last decade, much efforts have been paid on the individual CR species measurement. The knee of the light species at 3-5 PeV is observed by KASCADE-Grande \cite{KASCADE_light}, while the ARGO-YBJ and a Cherenkov telescope of LHAASO measure the H+He knee at 700 TeV\cite{2015PhRvD..92i2005B}. Meanwhile, a heavy knee (mainly by iron) is also published by the KASCADE-Grande at 80 PeV \citep{2011PhRvL.107q1104A}. Within the experimental uncertainties, we find both the light and heavy CR species share a common Lorentz factor at the order of $10^6$.
\par
Recently, the electron spectrum has been measured up to about 20 TeV \citep{2015arXiv150806597S, 2008Natur.456..362C, 2011ICRC....6...47B, 2009A&A...508..561A, hesstalk, dampe}. It can be described by a broken power law with an index about -3.1 before $\sim 1$ TeV and nearly -4.1 after $\sim 1$ TeV \citep{2015arXiv150806597S}. In the past years, much attention has been paid to the excesses of the electron and the positron including both the dark matter (\citep{2013FrPhy...8..794B} and references therein) and the astrophysics origin \citep{2004ApJ...601..340K, 2009PhRvL.103k1302S, 2009PhRvD..80l3017A, 2013ApJ...772...18L, 2010MNRAS.406L..25H, 2011PhRvD..83b3002K, 2012PhRvD..85d3507L, 2012MNRAS.421.3543K}. Several studies attribute this spectral break to the acceleration limit or the confinement at source \citep{2009A&A...497...17V, 2012MNRAS.427...91O, 2011MNRAS.414.1432T, 2018ApJ...854...57F}, the energy loss during propagation \citep{2010NJPh...12c3044S, 2010ApJ...710..236S}, or the contribution from nearby pulsars \citep{2010ApJ...710..958K}.
\par
The energy scale of the CR knee and the electron cutoff are quite different, which is reasonable to be explained with independent processes. But it should be noticed that both PeV CR nuclei and TeV electrons have almost a same Lorentz factor $\gamma$ around $10^6$. So it is possible that these similar spectral features share a common origin. This work will try to explore this possibility.

\section{Model Construction}
\subsection{Model Description}
Suppose that these spectral breaks with a same Lorentz factor does make a physical sense, a natural explanation is to associate it with a threshold interaction. The well-known GZK effect \citep{1988A&A...199....1B} is a threshold interaction $\gamma \rightarrow \pi^0, \ \pi^{+}$ and predicts CR's $\sim$ 50-EeV cutoff. This prediction is consistent with the observation from HiRes \cite{GZK_HiRes}, the Pierre Auger Observatory \cite{GZK_Auger} and the Telescope Array \cite{GZK_TA}. Similar threshold process $\gamma \rightarrow e^+, \ e^-$ can also be used to explain the spectral knee \citep{2009ApJ...700L.170H, 2010SCPMA..53..842W}. However, the $e^{+} e^{-}$ pair production process is not sufficient if it occurs in the whole Galaxy. Considering its cross section with 1 eV photon is $\sim 10$ mb, together with the CR electron's lifetime $\sim 10^5$ years in Galaxy, at least $10^4 \ \rm cm^{-3}$  are required for one collision. In fact, the observed number density of $1$ eV photons is only $1 \ \rm cm^{-3}$. Hence, we propose that there exists a kind of new particle $X$ widespread in the Galaxy and it can interact with CRs as
\begin{equation}\label{interaction}
\rm CR \ + \ X \ \longrightarrow \ CR \ + \ X^{\prime}
\end{equation}
where $\rm X^{\prime}$ represents the product after collision and is much heavier than the $X$.
\par
The particle $X$ in the Galaxy can be either relativistic or non-relativistic, which has no influence on the spectral break. In order to account the minimal energy budget for the $X$'s production, it is set to be non-relativistic, i.e. $v_X = 0$ simply. Assuming the CR knee and the electron cutoff are not a recent phenomenon, the X particle should have a very long life time, which is possibly at a cosmological level. Thus, according to the interaction in Eq. \ref{interaction}, the threshold of CR is derived as
\begin{equation}\label{thresh}
E_{\rm th} = \frac{m_{X^{\prime}}}{m_X}( m_{\rm CR} + \frac12 m_{\rm X^{\prime}} - \frac{m_{\rm X}^2}{2m_{\rm X^{\prime}}} )
\end{equation}
\par
In the Eq. \ref{thresh}, the Lorentz factor $\gamma$ at the threshold will approximate to $m_{X^{\prime}} / m_X$ when both $m_{X}$ and $m_{X^{\prime}}$ are much less than $m_{CR}$. By assuming the major composition at knee is Helium, the ratio $m_{X^{\prime}} / m_X$ can thus be evaluated as $10^6$. On the other hand, the interaction of Eq. \ref{thresh} does not consider the dissociation effect of CR nuclei. In the case $m_X < 1$ eV the collision energy will not exceed the CR nuclei's binding energy around the threshold. Therefore we leave it out of consideration in this work.
\par
Moreover, the non-threshold interactions between the CR and the $X$ will take place in the total energy range. Two typical processes includes the conversion into photons,
\begin{equation}
\label{gamma}
\rm CR \ + \ X \ \longrightarrow \ CR \ + \ \gamma
\end{equation}
and the elastic collision of the $X$.
\begin{equation}
\label{elastic}
\rm CR \ + \ X \ \longrightarrow \ CR \ + \ X
\end{equation}
\par
If both of them play the essential role, the effect of the interaction Eq. \ref{thresh} will be highly suppressed. Hence, some properties should be introduced to the $X$ to eliminate their effects. The interaction in Eq. \ref{gamma} will be forbidden if both the $X$ and the $X^{\prime}$ bear a same conserved quantum number. On the other hand, in the case that the $X^{\prime}$ has a much stronger coupling than the $X$, the interaction in Eq. \ref{elastic} will be less important. It should be noted such an coupling is similar to the axion model \citep{2008PhRvD..78k5012P, 2010PhRvD..82f5006D}, whose strength depends on the axion's mass. And as will be seen later, the $X$'s mass range covers the axion. Under these assumptions, the non-threshold interactions are not considered in the calculation.

\subsection{Calculation Algorithm}
Dynamic modeling is beyond the scope of this work, we only calculate the kinetic effect of this model by performing the Monte Carlo (MC) method. In the consideration for a simple model, cross sections are set to be constants $\sigma_e$ for the electron and $A^2 \sigma_P$ for the nuclei above the threshold energy, where A is the mass number and $A^2$ is used cosidering the coherence of the nucleons. Moreover, final particles after the interaction are assumed to have an isotropic angular distribution in the central-mass frame. Suppose that the $X$s may be coherent, the interaction rate can be enhanced and we introduce a enhanced factor $\eta$ on the cross section.
\par
Denote the number density of the $X$ in the galaxy by $n_x$, and the mean free time of the interaction by $\tau_A$ for the nuclei and $\tau_e$ for the electron. Then we can evaluate $\tau_A \ = \ 1 / n_x A^2 \eta \sigma_P c$, and $\tau_e \ = \ 1 / n_x \eta \sigma_e c$. The total life time of CRs confines the total interaction times before arriving at the observer. And it is estimated as $\tau_{esc}(R) \ \sim \ 2 \times 10^8 ( \frac{R}{1 \ GV} )^{-0.6}$ yr \citep{2009A&A...497..991P} for the nuclei and $10^5$ yr  \citep{2016arXiv160706615A} for the electron, given that we only concern the electron around 1 TeV. The escape time $\tau_{esc}(R)$ depends on the the Boron to Carbon flux ratio measurement. And AMS-02 fits the power index with -0.333 from 1.9 GV to 2.6 TV \citep{2015PhRvL.115u1101A}, which indicates $\tau_{esc}(R) \propto R^{-0.333}$. We find that such a change does not substantially modify our results, except lowering the fitting cross section by about one order of magnitude.
\par
As a common view, CR's propagation process in the Galaxy is regarded as a major magnetic-governed diffusion process \citep{2007ARNPS..57..285S}, where the secondary productions and energy loss processes can take place during the propagation. The diffusion process affects the observed spectrum by a soften index, which depends on the diffusion coefficient. And the energy-loss processes such as Inverse Compton scattering and synchrotron radiation has a severe impact on the observed electron spectrum. Actually, as measured by AMS-02 \citep{2014PhRvL.113v1102A}, the $e^+ + e^-$ spectrum presents a well single power law with the index $\sim - 3.17$ before 800 GeV. Hence, we can deal with the propagation effect and the interaction with the $X$ separately for simplicity. In the calculation, the single power-law spectra is injected primarily as the representative of the spectra after the CR's transport.
\par
There are three free parameters need to be adjusted, including the $X$'s mass $m_x$, the interaction parameters $\eta n_x \sigma_P$ for the proton, and $\eta n_x \sigma_e$ for the electron. Detail information of the calculation and the results are shown in the next section.

\section{Calculations and Results}
\subsection{All Particle Spectrum}
In the calculation of the CR nuclei, major species from the P to Fe are injected into an $X$-abundant space with the spectra suggested by Horandel \citep{2003APh....19..193H}. Under the fact that $m_{X^{\prime}} \ll m_{CR}$, we find that the final nuclei spectra show no dependence on the input mass $m_X$. So a representative mass 1 eV is calculated as an illustration and the parameter $\eta n_x \sigma_P$ is tuned as $4 \times 10^{-21} \ cm^{-1}$ to explain the data.
\par
\begin{center}
\begin{figure*}
\centering
\includegraphics[width=.45\textwidth]{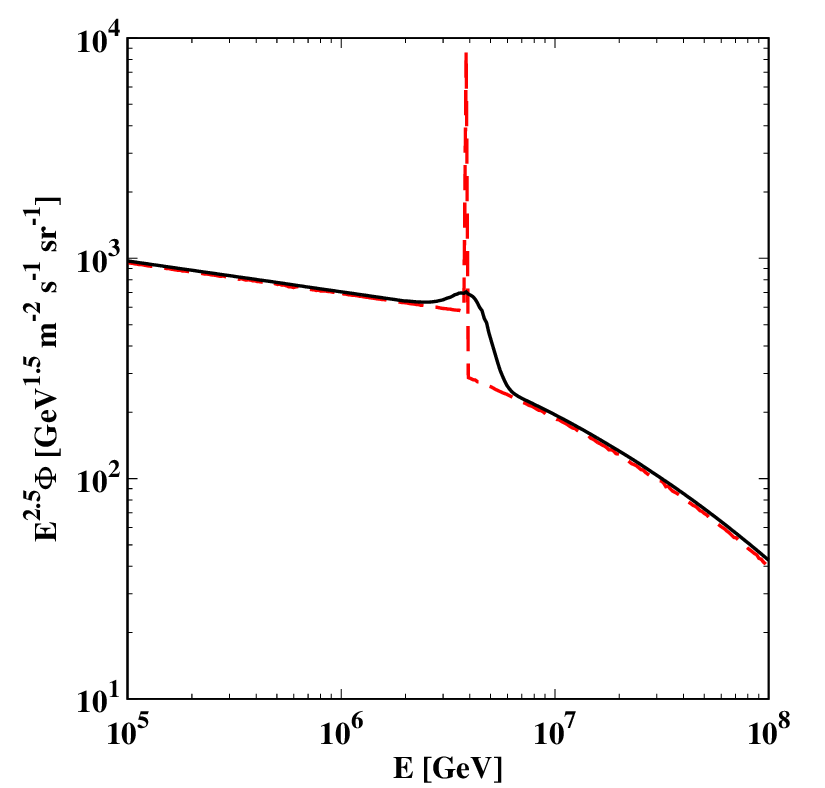}
\includegraphics[width=.45\textwidth]{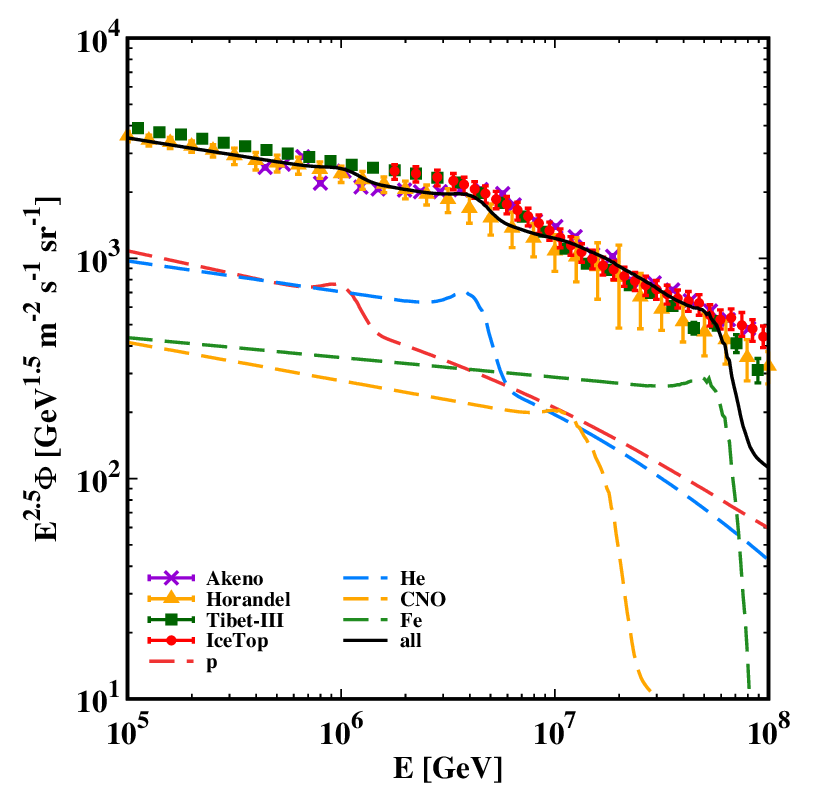}
\caption{Left: the spectra of helium. The red solid line is the spectrum involving the interaction with particle X, while the black line is the same one but taking the energy resolution of the detector into account, which is set to $20\%$ here. Right: The spectra of all-particle (black) and each component (including P(red), He(blue), CNO(yellow) and Fe(green) ), considering $20\%$ energy resolution. The data points are taken from Akeno \cite{1993PhRvD..48.1949I}(violet cross), Horandel \cite{2003APh....19..193H}(yellow triangle), Tibet-III \cite{2008ApJ...678.1165A}(green square) and IceTop \cite{2013PhRvD..88d2004A}(red circle) respectively.}
\label{fig:nuclei}
\end{figure*}
\end{center}

The direct MC calculation result for He is illustrated in the left panel of Fig. \ref{fig:nuclei} with the red dashed line and presents a sharp peak at the threshold. This peak is caused by the pile-up effect from the lost particles at higher energy originally. It can be derived that its average energy loss $<\triangle E>$ is about $\gamma^2 m_X \sim$ TeV at the threshold. Accordingly, the width of the peak is at the order of TeV.
\par
On the other hand, considering the actual energy resolution for experiments, we should not expect to observe the peak anyway. If the energy resolution cab can be described by a gaussian distribution, the observed spectra through the formula can be obtained by the following integration
\begin{equation}\label{eq:en_res}
F_{obs}(E) = \int F_{true}( E_0 ) \ \frac{1}{\sqrt{2 \pi} \sigma_{res}} exp \left[ - \frac{ (E - E_0)^2 }{ 2 \sigma_{res}^2} \right] d E_0
\end{equation}
Where the parameter $\sigma_{res}$ denotes the energy resolution with an approximate value $20\% E_0$ standing for a typical experiment, variable $E_0$ denote the true energy while E the observed energy, and $F_{true}$ the true flux. The expected observation for the Helium is illustrated in the left panel of Fig. \ref{fig:nuclei} with black solid line, and its peak is thus wiped off.
\par
The right panel of Fig. \ref{fig:nuclei} illustrates the spectra of all-particle and the main element, including P, He, CNO, and Fe. It can be seen that the all-particle spectrum agrees well with the observation below $60$ PeV. Moreover, the zig-zag shape appears at the threshold for each nuclei specie, which is a typical characteristic of the threshold interaction. Such an interaction causes a spectral deformation due to the energy loss, which can be estimated simply
\begin{equation}\label{observation}
N(E) \propto \begin{cases}
E^{-\alpha}, \ E < E_{th} \\
(E+\frac{\tau_{esc}}{\tau_A}<\triangle E>)^{-\alpha}, \ E > E_{th}
\end{cases}
\end{equation}
The parameter $\alpha$ denotes the injected spectral index. Noticed that $<\triangle E> \sim \gamma^2 m_X$, where $\gamma = E / m_{CR}$, it can be derived that the spectrum above the threshold will get softer than that below the threshold. Therefore, a knee-like feature is naturally formed. Besides, the parameter $\tau_{esc}$ is energy-dependent, and it has an impact on the observed spectrum as well.

\subsection{Electron Spectrum}
\begin{center}
\begin{figure*}
\centering
\includegraphics[width=.45\textwidth]{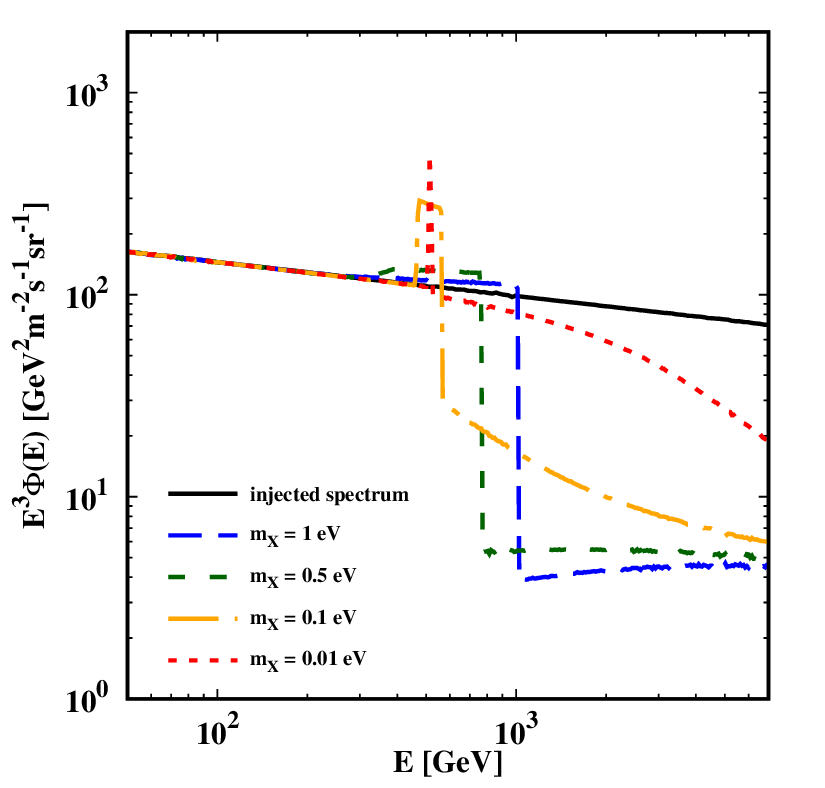}
\caption{The electron spectra after interacting with particle X of different masses. The black solid line is the original spectrum, while the other lines correspond to the Monte Carlo calculation with different X masses $m_{\rm X}$ respectively, i.e. $1$ eV(blue dash), $0.5$ eV(green short-dash), $0.1$ eV(yellow dash-dot), $0.01$ eV(red dot).}
\label{fig:electron_vary}
\end{figure*}
\end{center}

Considering that the electron mass is much smaller than the CR nuclei, the calculated break position of the electron will be sensitive to the $m_X$. In the Fig. \ref{fig:electron_vary}, we plot the calculated electron spectra with a set of $m_X$ from 1 eV to 0.01 eV. As the $m_X$ decreases, the break position varies from 1 TeV to 500 GeV according to Eq. \ref{thresh}, together with a gradually prominent peak. When $m_X \ll m_e$, the calculated electron spectrum will behave similar to the CR nuclei accordingly.
\par
\begin{center}
\begin{figure*}
\centering
\includegraphics[width=.45\textwidth]{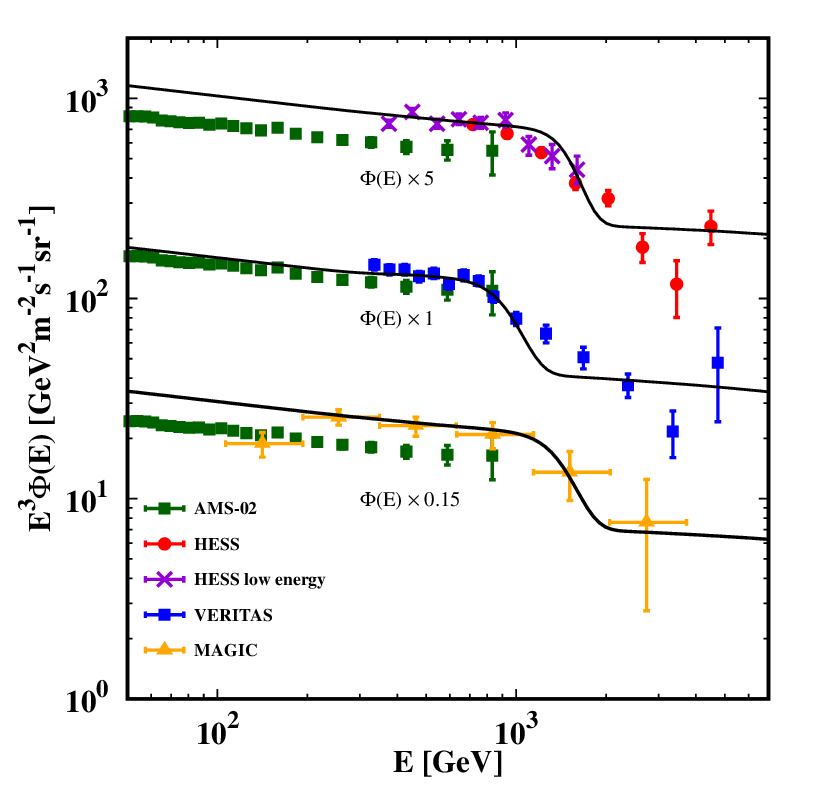}
\includegraphics[width=.45\textwidth]{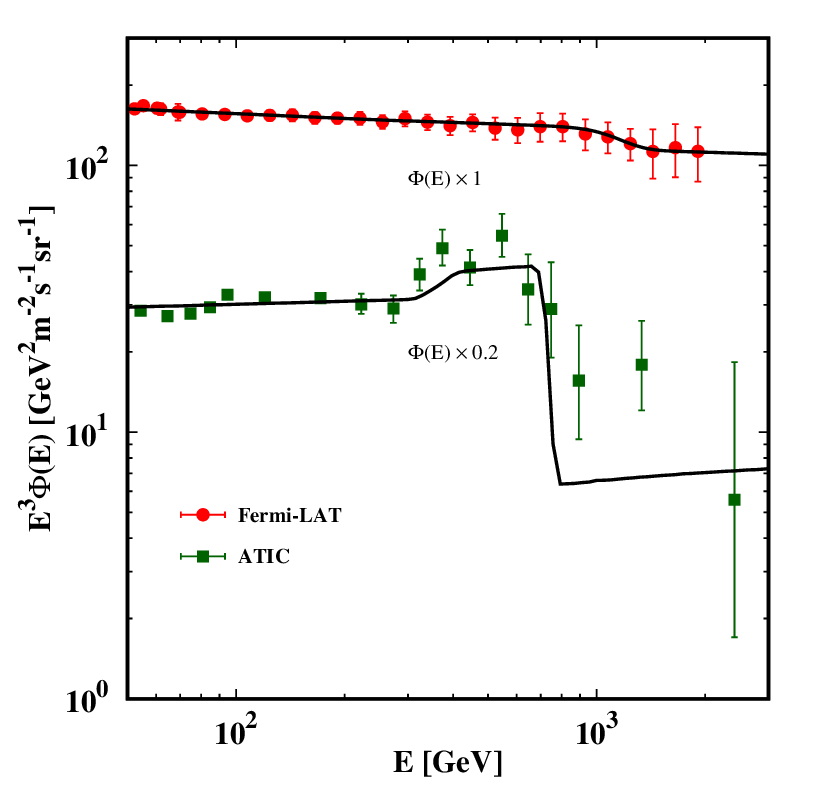}
\caption{Left: The fitting to the observational results of early HESS \cite{2008PhRvL.101z1104A, 2009A&A...508..561A}, VERITAS \cite{2015arXiv151001269S} and MAGIC \cite{2011ICRC....6...47B}, where the black solid line is the fitting to different observational data, given the energy resolution and energy rescale. Right: The fitting to the measurements of Fermi-LAT \cite{2017arXiv170301073F} and ATIC \cite{2008Natur.456..362C}, where the black solid line is the fitting to different observational data, given the energy resolution. The energy resolutions for each experiment are $15\%$ for HESS \cite{2009A&A...508..561A}, $20\%$ for VERITAS \cite{2015arXiv151001269S}, $20\%$ for MAGIC \cite{2011ICRC....6...47B}, $2\%$ for ATIC \cite{2008Natur.456..362C}, and $17\%$ for Fermi \cite{2017arXiv170301073F} respectively.  }
\label{fig:electron}
\end{figure*}
\end{center}

Next, a more realistic calculation for the electron is implemented to explain the experiments. Considering the high precision electron spectrum from AMS-02 \citep{2014arXiv1402.0437C}, We adopt the primarily injecting electron spectrum by fitting to the AMS02's result between 30 GeV and 300 GeV in an index $\sim -3.17$. And the effect of the energy resolution is also considered as claimed by other experiments. Noticed that there are differences about the absolute flux between the AMS-02 with others, it is considered that they are caused by different energy scale for each experiment. Then, we treat the AMS-02's energy scale as the standard one, and use parameter $\xi$ to describe the relative energy scale. The calculation results of  early HESS \cite{2008PhRvL.101z1104A, 2009A&A...508..561A}, VERITAS \cite{2015arXiv151001269S} and MAGIC \cite{2011ICRC....6...47B} are shown in the left panel in Fig. \ref{fig:electron}.
And the calculation results agree well with each experiment, together with the zig-zag feature as shown in the CR nuclei.
\par
Different from the ground-base experiments, both Fermi-ALT \cite{2017arXiv170301073F} and ATIC \cite{2008Natur.456..362C} have low energy observations and are harder than the AMS-02. So we just use their lower energy part as injection spectrum and results are shown in the right panel in Fig. \ref{fig:electron}. The fitted parameters of all the experiments are listed in Tab. \ref{tab:1}. Compared with other experiments, the measurement of Fermi-LAT shows a less prominent structure, and the fitted cross section $\eta n_{\rm x} \sigma_e$ is one order of magnitude smaller. Meanwhile the bump feature of ATIC experiment can be reproduced when $m_x \sim 0.4$ eV. Thus, it can be seen that the $m_x$ is very sensitive to the experiments. Results spreading from 0.4 eV to 1.6 eV indicates that the experimental uncertainty of the $m_x$ is in an order of 1 eV. Namely, the true $m_x$ can be any number from 0 to about 1 eV.

\begin{table}
\centering
\caption{\label{tab:1} The fitting parameters.}
\begin{ruledtabular}
\begin{tabular}{ccccccc}
  \toprule
  parameters                & HESS & VERITAS & MAGIC & ATIC & Fermi \\
  \hline
  $m_x$ (eV)                & 1.6 & 0.7 & 1.4 & 0.4 & 1.1  \\
  $\eta n_x \sigma_e$  & 1.2 & 1.2 & 1.2 & 2.4 & 0.2  \\
  ($\times 10^{-23}$ cm$^{-1}$) \\
  $\xi$                           & 1.08 & 1.01 & 1.09 & 1 & 1  \\
  \bottomrule
\end{tabular}
\end{ruledtabular}
\end{table}

\subsection{Modified Cross Section}
\begin{center}
\begin{figure*}
\centering
\includegraphics[width=.45\textwidth]{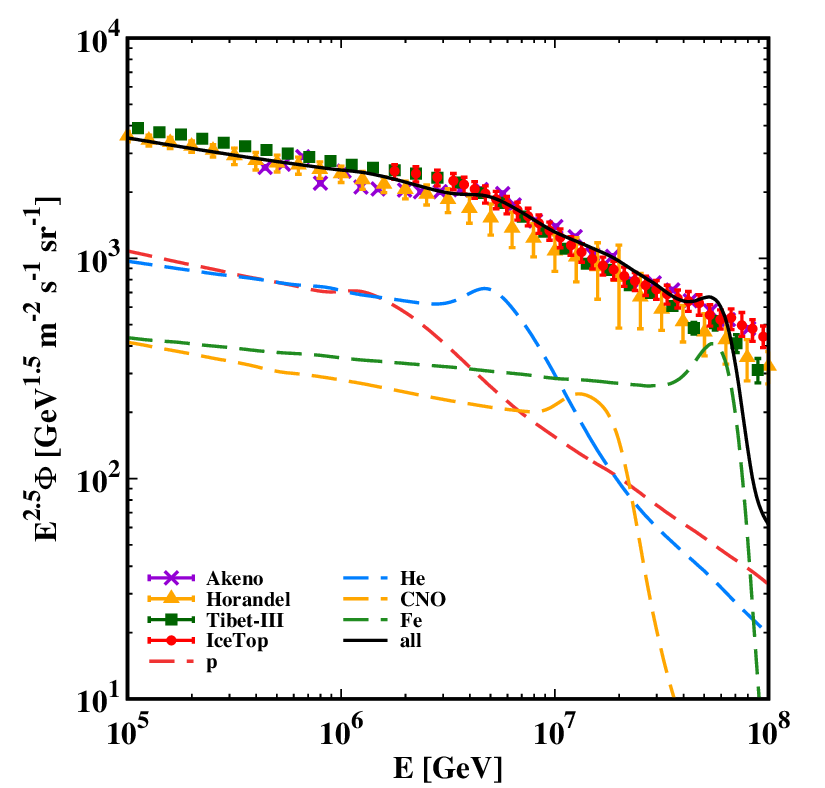}
\includegraphics[width=.45\textwidth]{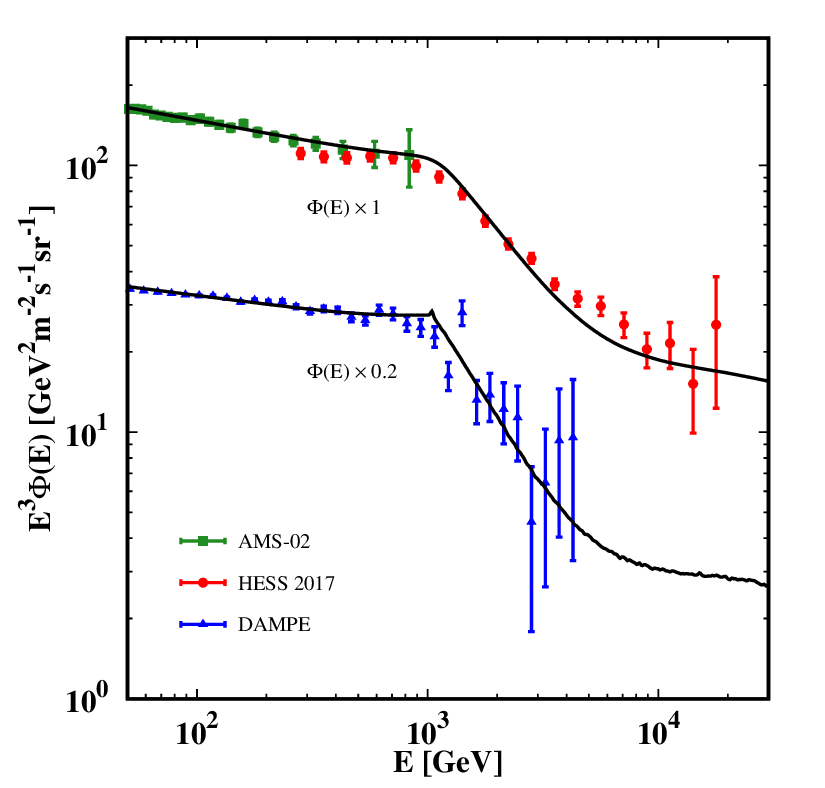}
\caption{The fit to the CR knee (left) and the latest electron spectra by HESS \cite{hesstalk} and DAMPE \cite{dampe} (right) with improve cross section.}
\label{fig:improvedcs}
\end{figure*}
\end{center}

Recently, both HESS \cite{hesstalk} and DAMPE \cite{dampe} experiments publish their observations of electron spectrum at TeV energies respectively. Compared with the early result, the latest measurement of HESS shares a much higher statistics. The former simplified cross section is difficult to explain the latest spectrum. Moreover, according to the precise result of DAMPE satellite, the spectral break occurs at $\sim 900$ GeV. To accommodate with these latest observations, an energy-dependent cross section is adopted which rises across the threshold and asymptotically approaches to a constant. It takes the form as
\begin{equation}
\sigma(E) = \left\{
\begin{aligned}
& 0 ~, & E < E_{th} \\
& \sigma_0 \left(1 -\exp \left[-\dfrac{(E-E_{th})}{2E_{th}} \right] \right) ~, & E > E_{th}
\end{aligned}
\right.
\end{equation}
Where the parameter $\sigma_0$ is the constant cross section as $\eta \sigma_{P}$ for the proton and $\eta \sigma_{e}$ for the electron. The results are shown in the left of Fig. \ref{fig:improvedcs} and the fitted parameters are listed in Tab. \ref{tab:2}.
\par
To test the effect of the energy-dependent cross section, we re-compute the knee likewise, where the parameter $\eta n_{x} \sigma_{P}$ is fitted as $6 \times 10^{-21}$ cm$^{-1}$. And the result is shown in the right of Fig. \ref{fig:improvedcs}. Compared with the formerly step-function cross section, the new calculation describes the CR knee and electron's cutoff equally well.

\begin{table}
\centering
\caption{\label{tab:2} The fitting parameters of HESS 2017 and DAMPE.}
\begin{ruledtabular}
\begin{tabular}{ccccccc}
  \toprule
  parameters                & HESS 2017 & DAMPE \\
  \hline
  $m_x$ ( eV )              & 1.07 & 1.07 \\
  $\eta n_x \sigma_e$  & 2 & 2.8 \\
  ( $\times 10^{-23} \ \rm cm^{-1}$ )\\
  $\xi$                           & 1 & 1 \\
  \bottomrule
\end{tabular}
\end{ruledtabular}
\end{table}

\section{Conclusions And Discussions}
Notice that the CR knee and electron's TeV spectral break may share a common Lorentz factor, we introduce a new threshold interaction with an unknown particle X during CRs' transport. Along with a simple model assumption, both of these spectral breaks can be well reproduced. Moreover, our model predicts the A-dependent knee of the CR nuclei, not Z-dependent, which is an essential characteristic for the future test on the spectral measurement.
\par
The major uncertainty of this model is the value of the common Lorentz factor. Latest measurement from ARGO \citep{2015PhRvD..92i2005B} and KASCADE-Grande \cite{2011PhRvL.107q1104A} indicates it ranges from $0.7 \times 10^6$ to $1.4 \times 10^6$ with the energy resolution around 25\%. Latest measurements of the electron reveal this value $1.4-1.8 \times 10^6$. In the case that the Lorentz factor are precisely equal for both the CR nuclei and the electron, the mass of the $X$ is allowable for an ultra light one, which covers the estimation of the axion \cite{2008PhRvD..78k5012P, 2010PhRvD..82f5006D, 2009PhRvL.103k1301S}, axion-like particle \cite{2010ARNPS..60..405J, 2013arXiv1311.0029E} and fuzzy dark matter (DM) \cite{fuzzydarkmatter}. So precise measurement of spectra for all components crossing their knee/cutoff region is important.
\par
Considering the high abundance of the Xs in the Galaxy, it is probable to regard the X particles as the major DM candidate, whose local energy density is evaluated as $\sim 0.4 \ \rm GeV/cm^3$ \cite{2015JCAP...12..001P}. Accordingly, the enhanced cross sections are derived $\sim 10 \ \mu$b for protons and $\sim 100$ nb for electrons. Latest measurement from EDGES \cite{nature1, nature2} also suggest a much larger cross section $\sim$ 1000 b for the DM and baryons scattering.
\par
As mentioned above, the particle X is allowable for the ultra light DM. Current detections for the weak-interaction massive particle, whose mass is expected about 100 GeV, will not bring constraints on our model. Besides, much efforts have been paid in searching for the light DM axion, and their mainly searching channels are through the axion couples with the electromagnetic field \cite{axion1, axion2, axion3}. Noted that a new conserved quantum number is assumed on the particle X, those channels are forbidden. An exotic feature of the axions is that they are proposed to form a condensate with long-range correlation \cite{2009PhRvL.103k1301S, axioncorrelation}. As the momentum transfer q is ultra low, the number of the coherent Xs within the corresponding volume is large and the cross section is enhanced by $\eta \sim N^2$, where N is the total X number in this volume. It is similar to the coherent neutrino-nucleus scattering process \cite{neutrino}. In the case that the X particle is at the scale of the fuzzy DM, i.e. $m_x \sim 10^{-22}$ eV, q is evaluated about $10^{-10}$ eV and the corresponding region contains N $\sim 10^{46}$ X particles. Therefore, any incoherent processes are beyond the current detecting ability, such as the X particles produced on the collider. The threshold requirement PeV for baryons and TeV for electrons also limit the observation of both the particle physics and astronomical experiments.
\par
Furthermore, whether all CR components have a same Lorentz factor is still questionable. Even if the mechanism of the knee is different from the electron's cutoff, our model is still viable assuming the X particle interacts only with the CR nuclei. Thus measurement about whether the spectral breaks of individual CR species are A-dependent is essential. Nevertheless, the detection for the high-energy CRs provides a unique way to explore the interactions with the ultra-light DM particles.

\textit{Acknowledgements.---}  We thank Xiaojun Bi, Shouhua Zhu, Chun Liu, Pengfei Yin and Yu Gao for their helpful discussions. This work is supported by the Natural Sciences Foundation of China (11135010, 11635011, 11761141001) and the National Key Program for Research and Development (2016YFA0400200).


\begin{thebibliography}{99}


\bibitem{1959JETP...35....8K}
G. V. {Kulikov}, G. B. {Khristiansen}, SJETP, {\bf 35}, 8 (1959)

\bibitem{1999JETP...89..391B}
E. G. {Berezhko}, L. T. {Ksenofontov}, SJETP, {\bf 89}, 391 (1999)

\bibitem{2002PhRvD..66h3004K}
K. {Kobayakawa}, Y. S. {Honda}, and T. {Samura}, Phys. Rev. D, {\bf 66}, 083004 (2002)

\bibitem{2001A&A...369..269B}
P. L. {Biermann}, N. {Langer}, E. S. {Seo} et al, A\&A, {\bf 369}, 269 (2001)

\bibitem{2003A&A...409..799S}
L. G. {Sveshnikova}, A\&A, {\bf 409}, 799 (2003)


\bibitem{1993A&A...268..726P}
V. S. {Ptuskin}, et al., A\&A, {\bf 268}, 726 (1993)

\bibitem{1998A&A...330..389W}
B. {Wiebel-Sooth}, P. L. {Biermann}, and H. {Meyer}, A\&A, {\bf 330}, 389 (1998)

\bibitem{2009ApJ...700L.170H}
H. B. Hu et al, ApJ, {\bf 700}, L170 (2009)

\bibitem{2010SCPMA..53..842W}
B. Wang et al, SCPMA, {\bf 53}, 842 (2010)

\bibitem{KASCADE_light}
T. Antoni et al., Astropart. Phys. {\bf 24}, 1¨C25 (2005)

\bibitem{2015PhRvD..92i2005B}
B. {Bartoli}, P. {Bernardini}, X.~J. {Bi} et al, Phys. Rev. D, {\bf 92}, 092005 (2015)

\bibitem[Apel et al.(2011)]{2011PhRvL.107q1104A}
Apel, W.~D., Arteaga-Vel{\'a}zquez, J.~C., Bekk, K., et al, Phy. Rev. Lett., {\bf 107}, 171104 (2011)

\bibitem{2015arXiv150806597S}
D. {Staszak} et al (VERITAS Collaboration), arXiv:1508.06597

\bibitem{2008Natur.456..362C}
J. Chang et al, Nature, {\bf 456}, 362 (2008)

\bibitem{2011ICRC....6...47B}
D. {Borla Tridon}, ICRC, {\bf 6}, 47 (2011)

\bibitem{2009A&A...508..561A}
F. {Aharonian} et al, A\&A, {\bf 508}, 561 (2009)

\bibitem{hesstalk}
https://indico.snu.ac.kr/indico/event/15/session/5
/contribution/694/material/slides/0.pdf, retrieved 13th July 2017

\bibitem{dampe}
DAMPE Collaboration, arXiv:1711.10981

\bibitem{2013FrPhy...8..794B}
X. J. Bi, P. F. Yin, Q. Yuan, Frontiers of Physics, {\bf 8}, 794 (2013)

\bibitem{2004ApJ...601..340K}
T. {Kobayashi}, Y. {Komori}, K. {Yoshida}, and J. {Nishimura}, ApJ, {\bf 601}, 340 (2004)

\bibitem{2009PhRvL.103k1302S}
N. J. Shaviv, E. Nakar, T. Piran, Phys. Rev. Lett., {\bf 103}, 111302 (2009)

\bibitem{2009PhRvD..80l3017A}
M. Ahlers, P. Mertsch, S. Sarkar, Phys. Rev. D, {\bf 80}, 123017 (2009)

\bibitem{2013ApJ...772...18L}
T. {Linden}, and S. {Profumo}, ApJ, {\bf 772}, 18 (2013)

\bibitem{2010MNRAS.406L..25H}
J. S. Heyl, R. Gill, L. Hernquist, MNRAS, {\bf 406}, L25 (2010)

\bibitem{2011PhRvD..83b3002K}
K. Kashiyama, K. Ioka, N. Kawanaka, Phys. Rev. D, {\bf 83}, 023002 (2011)

\bibitem{2012PhRvD..85d3507L}
J. {Liu}, Q. {Yuan}, X.-J. {Bi} et al, Phys. Rev. D, {\bf 85}, 043507 (2012)

\bibitem{2012MNRAS.421.3543K}
S. Kisaka, \& N. Kawanaka, MNRAS, {\bf 421}, 3543 (2012)

\bibitem{2009A&A...497...17V}
G. {Vannoni}, S. {Gabici}, and F.~A. {Aharonian}, A\&A, {\bf 497}, 17 (2009)

\bibitem{2012MNRAS.427...91O}
Y. {Ohira}, R. {Yamazaki}, N. {Kawanaka}, and K. {Ioka}, MNRAS, {\bf 427}, 91 (2012)

\bibitem{2011MNRAS.414.1432T}
S. {Thoudam}, and J. R. {H{\"o}randel}, MNRAS, {\bf 414}, 1432 (2011)

\bibitem{2018ApJ...854...57F}
K. {Fang}, X.-J. {Bi} and P.-F. {Yin}, APJ, {\bf 854}, 57 (2018)

\bibitem{2010NJPh...12c3044S}
R. {Schlickeiser}, and J. {Ruppel}, New Journal of Physics, {\bf 12}, 033044 (2009)

\bibitem{2010ApJ...710..236S}
{\L}. {Stawarz}, V. {Petrosian}, and R.~D. {Blandford}, ApJ, {\bf 710}, 236 (2010)

\bibitem{2010ApJ...710..958K}
N. {Kawanaka}, K. {Ioka}, and M.~M. {Nojiri}, ApJ, {\bf 710}, 958 (2010)

\bibitem{1988A&A...199....1B}
V.~S. {Berezinskii} and S.~I. {Grigor'eva}, A\&A, {\bf 199}, 1 (1988)

\bibitem{GZK_HiRes}
P. Sokolsky and G. B. Thomson, J. Phys. G: Nucl. Part. Phys. {\bf 34}, R401¨CR429 (2007)

\bibitem{GZK_Auger}
R. Caruso and Pierre Auger Collaboration, Astrophys. Space Sci. Trans., {\bf 7}, 445-451 (2011)

\bibitem{GZK_TA}
T. Abu-Zayyad et al., Astropart. Phys. {\bf 61}, 93 (2015) 

\bibitem{2008PhRvD..78k5012P}
M. {Pospelov}, A. {Ritz} and M. {Voloshin}, PRD, {\bf 78}, 115012 (2008)

\bibitem{2010PhRvD..82f5006D}
A. {Derevianko}, V.~A. {Dzuba}, V.~V. {Flambaum} and M. {Pospelov}, PRD, {\bf 82}, 065006 (2010)

\bibitem{2009A&A...497..991P}
A. {Putze}, et al, A\&A, {\bf 497}, 991 (2009)

\bibitem{2016arXiv160706615A}
R. {Attallah}, Probing the astrophysical origin of high-energy cosmic-ray electrons with Monte Carlo simulation, arXiv:1607.06615

\bibitem{2015PhRvL.115u1101A}
M. {Aguilar} et al, Phys. Rev. Lett., {\bf 117}, 231102 (2016)

\bibitem{2007ARNPS..57..285S}
A. W. {Strong}, I. V. {Moskalenko}, and V. S. {Ptuskin}, Annual Review of Nuclear and Particle Science, \textbf{57}, 285 (2007)

\bibitem{2014PhRvL.113v1102A}
M. {Aguilar}, et al, Phys. Rev. Lett., {\bf 113}, 221102 (2014)

\bibitem{2003APh....19..193H}
J. R. {H{\"o}randel}, Astroparticle Physics, {\bf 19}, 193 (2003)

\bibitem{1993PhRvD..48.1949I}
M. {Ichimura}, et al, Phys. Rev. D, {\bf 48}, 1949 (1993)

\bibitem{2008ApJ...678.1165A}
Amenomori, M., Bi, X.~J., Chen, D. et al, APJ, {\bf 678}, 1165-1179 (2008)

\bibitem[Aartsen et al.(2013)]{2013PhRvD..88d2004A}
Aartsen, M.~G., Abbasi, R., Abdou, Y. et al, PRD, {\bf 88}, 042004 (2013)

\bibitem{2014arXiv1402.0437C}
C. {Corti} and {for the AMS collaboration}, The cosmic ray electron and positron spectra measured by AMS-02, arXiv:1402.0437

\bibitem[Aharonian et al.(2008)]{2008PhRvL.101z1104A}
Aharonian, F., Akhperjanian, A.~G., Barres de Almeida, U. et al, Phy. Rev. Lett., {\bf 101}, 261104 (2008)

\bibitem{2015arXiv151001269S}
D. {Staszak}, et al, arXiv:1510.01269

\bibitem{2017arXiv170301073F}
{Fermi-LAT Collaboration}, arXiv:1703.01073

\bibitem{2009PhRvL.103k1301S}
P. {Sikivie} and Q. {Yang}, Phys. Rev. Lett., {\bf 103}, 111301 (2009)

\bibitem{2010ARNPS..60..405J}
T. {Jaeckel} and A. {Ringwald}, Ann. Rev. Nucl. Part. Sci. {\bf 60}, 405 (2010)

\bibitem{2013arXiv1311.0029E}
R. {Essig} et al, arxiv:1311.0029

\bibitem{fuzzydarkmatter}
W. Hu, R. Barkana, and A. Gruzinov, Phys. Rev. Lett., {\bf 85}, 1158 (2000)






\bibitem{2015JCAP...12..001P}
M. {Pato}, F. {Iocco} and G. {Bertone}, JCAP, {\bf 12}, 001 (2015)

\bibitem{nature1}
Rennan Barkana1, Nature, {\bf 555}, 71 (2018)

\bibitem{nature2}
Judd D. Bowman, Alan E. E. Rogers, Raul A. Monsalve et al, Nature, {\bf 555}, 67 (2018)

\bibitem{axion1}
S. Andriamonje et al, JCAP, {\bf 4}, 010 (2007)

\bibitem{axion2}
J. Hoskins et al, Phys. Rev. D, {\bf 84}, 121302 (2011)

\bibitem{axion3}
S. J. Asztalos et al, Phys. Rev. Lett., {\bf 104}, 041301 (2010)

\bibitem{axioncorrelation}
O. Erken et al, Phys. Rev. D, {\bf 85}, 063520 (2012)

\bibitem{neutrino}
Daniel Z. Freedman, Phys. Rev. D, {\bf 9}, 1389 (1974)

\end{thebibliography}
\end{document}